%% file: main_revised.tex
\documentclass[12pt,a4paper]{article}
\usepackage{arxiv}

\usepackage[utf8]{inputenc}
\usepackage[english]{babel}
\usepackage{amsmath,amssymb}
\usepackage{graphicx}
\usepackage{setspace}
\usepackage{hyperref}
\usepackage{natbib}
\usepackage{booktabs}
\usepackage{caption}

\hypersetup{
    colorlinks=true,
    linkcolor=blue,
    citecolor=blue,
    urlcolor=blue
}

\title{\textbf{Compound Deception in Elite Peer Review:\\A Failure Mode Taxonomy of 100\\Fabricated Citations at NeurIPS 2025}}

\author{
	Samar Ansari\\
	School of Computing and Engineering Sciences\\
	University of Chester\\
	Chester, CH1 4BJ, United Kingdom \\
	\texttt{m.ansari@chester.ac.uk}
}

\begin{document}

\maketitle

\begin{abstract}
Large language models (LLMs) are increasingly used in academic writing workflows, yet they frequently \textit{hallucinate} by generating citations to sources that do not exist. This study analyzes 100 AI-generated hallucinated citations that appeared in papers accepted by the 2025 Conference on Neural Information Processing Systems (NeurIPS), one of the world's most prestigious AI conferences. Despite review by 3--5 expert researchers per paper, these fabricated citations evaded detection, appearing in 53 published papers ($\approx$1\% of all accepted papers). We develop a five-category taxonomy that classifies hallucinations by their failure mode: Total Fabrication (66\%), Partial Attribute Corruption (27\%), Identifier Hijacking (4\%), Placeholder Hallucination (2\%), and Semantic Hallucination (1\%). Our analysis reveals a critical finding: every hallucination (100\%) exhibited compound failure modes. The distribution of secondary characteristics was dominated by Semantic Hallucination (63\%) and Identifier Hijacking (29\%), which often appeared alongside Total Fabrication to create a veneer of plausibility and false verifiability. These compound structures exploit multiple verification heuristics simultaneously, explaining why peer review fails to detect them. The distribution exhibits a bimodal pattern: 92\% of contaminated papers contain 1--2 hallucinations (minimal AI use) while 8\% contain 4--13 hallucinations (heavy reliance). These findings demonstrate that current peer review processes do not include effective citation verification and that the problem extends beyond NeurIPS to other major conferences, government reports, and professional consulting. We propose mandatory automated citation verification at submission as an implementable solution to prevent fabricated citations from becoming normalized in scientific literature.
\end{abstract}


\input{introduction_revised.tex}
\input{methods_revised.tex}
\input{results_revised.tex}
\input{discussion_revised.tex}
\input{conclusion_revised.tex}

\section*{Acknowledgments}
The author thanks the GPTZero team for making their NeurIPS 2025 hallucination analysis publicly available.

\bibliographystyle{unsrt}
\bibliography{references_updated}

\end{document}

%% file: introduction_revised.tex
\section{Introduction}

In December 2025, the Conference on Neural Information Processing Systems (NeurIPS), one of the world's most prestigious artificial intelligence conferences, was conducted with 5,290 accepted papers (reflecting a 24.52\% acceptance rate out of 21,575  submissions\footnote{https://blog.neurips.cc/2025/09/30/reflections-on-the-2025-review-process-from-the-program-committee-chairs/}). Each paper had undergone peer review by 3--5 expert researchers selected for their domain expertise. Yet recent analysis reveals that at least 53 of these papers ($\approx$1\%) contained fabricated citations that evaded detection throughout the review process \cite{gptzero2026neurips}. These are not typographical errors or formatting mistakes; they are citations to sources that do not exist, with invented authors and/or fabricated titles, and/or false publication details.

The appearance of AI-generated hallucinated citations in peer-reviewed literature represents a fundamental challenge to the integrity of scientific discourse. Citations serve as the evidentiary foundation of scholarly work, establishing what prior research has demonstrated, enabling reproducibility, and situating new contributions within existing knowledge. When citations are fabricated, these epistemic functions collapse. Readers cannot verify claims attributed to non-existent sources. Future researchers waste time searching for papers that were never published. The citation graph that structures scientific knowledge becomes contaminated with false linkages.

This problem extends beyond NeurIPS. Analysis of ICLR 2026 submissions found over 50 hallucinations in papers under review, some rated 8/10 by reviewers \cite{gptzero2025iclr}. Fabricated citations have also appeared in U.S. government reports requiring corrections \cite{weber_maha_2025} and professional consulting outputs requiring \$98,000 (AUD) refunds \cite{gptzero2025deloitte}. The common factor across these cases is the use of large language models (LLMs) in writing workflows combined with inadequate verification of AI-generated content.

LLMs are known to \textit{hallucinate\footnote{\textbf{For non-technical readers:} Think of an LLM like a hyper-advanced version of a smartphone’s autocomplete. It doesn’t actually \textit{know} facts; it just knows which words usually follow each other based on patterns it saw in the past. A \textit{hallucination} happens when the AI prioritizes sounding fluent and convincing over being accurate. Because its only goal is to finish the sentence in a way that sounds human, it will occasionally `fill in the blanks' with a made-up name or date simply because that word fits the mathematical rhythm of the sentence perfectly.}\textsuperscript{,}\footnote{\textbf{For technical readers:} From a technical standpoint, a hallucination is a failure of grounding, where the model’s stochastic output diverges from the training distribution or the provided context. During the decoding process, the model may sample a token that has high linguistic probability but zero factual density, leading to a `drift' where subsequent tokens are generated to maintain the internal logical consistency of the initial error. This is often exacerbated by exposure bias or high temperature settings, where the model prioritizes the entropy of the output over the objective constraints of the latent space.}}, generating fluent, confident-sounding text that is factually incorrect \cite{ji2023hallucination,huang2025survey}. Citations are particularly vulnerable to hallucination because they involve precise metadata (author names, titles, venues, publication dates, DOIs) that language models may blend, paraphrase, or fabricate while maintaining surface-level plausibility \cite{alkaissi2023artificial}. A model trained on millions of academic papers can generate citations that look professionally formatted, use appropriate terminology, and cite plausible venues, while corresponding to no actual publication.

The NeurIPS 2025 contamination is especially concerning because it occurred at a premier AI conference where authors and reviewers are experts in the very systems that produce hallucinations. If researchers who specialize in LLMs cannot reliably detect AI-generated fabrications in their own review process, the problem is systematic, not isolated.

\subsection{Current Understanding and Research Gaps}

Previous research has documented LLM hallucination in various contexts, including medical information \cite{lee2023benefits}, legal reasoning \cite{dahl2024large}, and general factual queries \cite{rawte2023survey}. The phenomenon extends beyond isolated hallucinations to encompass broader patterns of synthetic content degradation, including what has been termed \textit{AI Slop}, low-quality, derivative outputs that contaminate digital ecosystems and degrade informational value \cite{ansari2025slop}. Studies of citation behavior have identified problems with citation accuracy \cite{rivkin2020manuscript} and misconduct \cite{fang2011misconduct}. However, the intersection of these issues: AI-generated fabricated citations in peer-reviewed literature, has received limited systematic analysis.

Recent cases have been documented journalistically (e.g., the Deloitte refund, the MAHA report corrections) but not analyzed systematically to understand the failure modes that allow hallucinated citations to succeed. Simply knowing that ``AI generates fake citations'' does not explain why sophisticated peer review fails to detect them. Are all hallucinations equally detectable? Do certain types exploit specific weaknesses in reviewer behavior? What distinguishes citations that pass review from those that get caught?

\subsection{Study Contribution}

This study addresses these questions through a systematic failure mode analysis of 100 hallucinated citations that appeared in NeurIPS 2025 accepted papers. Rather than simply documenting that citations are ``fake,'' we developed a taxonomy that classifies hallucinations by their mechanism: how they deviate from legitimate scholarly practice and how these deviations succeed at evading detection.

Our analysis reveals that AI-generated hallucinations are predominantly Total Fabrications (66\%), with citations invented wholesale rather than corrupted from real sources. However, a critical finding emerged: every hallucination in our dataset exhibited compound failure modes, employing multiple deception strategies simultaneously. The distribution of secondary failure characteristics was dominated by Semantic Hallucination (63\% of all citations) and Identifier Hijacking (29\% of all citations), which often appeared alongside Total Fabrication to create a veneer of plausibility and false verifiability through working links to unrelated papers.

These findings have immediate implications for research integrity policy. The problem is not hypothetical or future-facing; it is documented in the published proceedings of a top-tier conference. The 2026 conference cycle represents an opportunity to implement verification requirements before fabricated citations become normalized in scientific literature.

The remainder of this paper is structured as follows: Section 2 describes our data source and analytical framework, including the five-category failure mode taxonomy. Section 3 presents the prevalence of each hallucination type, provides representative examples, and analyzes compound failure patterns. Section 4 analyzes why peer review failed to detect these fabrications, addresses counterarguments about the severity of citation errors, and proposes concrete solutions. Section 5 concludes with implications for research integrity and recommendations for conference organizers.

%% file: methods_revised.tex
\section{Methods}

\subsection{Data Source}

We obtained citation-level data from GPTZero's systematic analysis of papers accepted by the 2025 Conference on Neural Information Processing Systems (NeurIPS). GPTZero\footnote{https://gptzero.me/}, a Canadian AI detection company, used their automated Hallucination Check tool to scan 4,841 (of the 5290 papers accepted) papers from NeurIPS 2025. This tool flags citations that cannot be verified through web search, academic databases (including Google Scholar, PubMed, arXiv, and CrossRef), and DOI/URL validation. Each flagged citation was subsequently verified by human experts at GPTZero to confirm it represented a probable hallucination rather than a legitimate archival source or temporary indexing gap.

GPTZero's analysis identified 100 hallucinated citations spanning 53 distinct papers\footnote{The GPTZero webpage also mentions the figure of 51 papers at one place, but the table containing the details of the halluicinations is for 53 papers as mentioned on their page.} (representing 1\% of all accepted papers). These papers had each undergone peer review by 3--5 expert reviewers per the standard NeurIPS process, yet the fabricated citations evaded detection. For our analysis, we used the complete dataset of 100 verified hallucinations as documented in GPTZero's public report \cite{gptzero2026neurips}, which included the full citation text, associated paper title, and diagnostic notes explaining why each citation was classified as a hallucination.

\subsection{Analytical Framework: Failure Mode Taxonomy}

To understand how AI-generated citations succeed at evading peer review, we developed a five-category taxonomy based on the hallucination mechanism. Rather than simply documenting that citations are ``fake,'' we classified each by its failure mode, the specific way in which the citation deviates from legitimate scholarly practice. This taxonomy distinguishes between citations that are entirely fabricated versus those that strategically corrupt or hijack real scholarly metadata.

The five categories are:

\begin{enumerate}
\item \textbf{Total Fabrication (TF)}: The citation has zero overlap with any real scholarly work. The authors, title, venue, and identifiers (DOI, arXiv ID, URL) are all fabricated. These represent complete invention by the language model.

\item \textbf{Partial Attribute Corruption (PAC)}: The citation blends real and fabricated elements. This includes: real authors attributed to papers they did not write; real paper titles with fabricated authors; correct venues with wrong publication years; or matching authors and venues with corrupted titles. These citations exploit pattern recognition, as reviewers may recognize familiar names or venues without checking all details.

\item \textbf{Identifier Hijacking (IH)}: The citation uses a valid scholarly identifier (arXiv ID, DOI, or URL) that points to a real paper, but the authors, title, or other metadata do not match the paper at that identifier. This creates false verifiability: a reviewer who clicks the link sees a real paper and may assume the citation is correct without verifying the metadata matches.

\item \textbf{Semantic Hallucination (SH)}: The citation invents a plausible-sounding title that fits the research domain conceptually but does not correspond to any real work. The title sounds professionally appropriate (e.g., ``A survey of model compression techniques'') but cannot be verified.

\item \textbf{Placeholder Hallucination (PH)}: The citation contains obvious generation failures such as template variables (``Firstname Lastname''), incomplete identifiers (``arXiv:2305.XXXX''), or placeholder text (``To appear''). These represent cases where the language model failed to complete its generation task.
\end{enumerate}

\subsection{Coding Procedure}

The author manually classified each of the 100 hallucinated citations by reviewing the full citation text, comparing it against the GPTZero verification notes, and assigning both a primary and secondary failure category. The primary code identifies the most prominent failure characteristic, i.e. the feature that would be most evident to a human reviewer attempting verification. The secondary code captures additional deception mechanisms present in the same citation, revealing the compound nature of AI-generated hallucinations.

For citations exhibiting characteristics of multiple categories, the primary classification was based on the dominant failure mode. For example, a citation with both fabricated authors and an incomplete arXiv ID was coded as PH (primary) if the placeholder was the most obvious indicator of AI generation, with TF as secondary to capture the fabricated authorship.

The coding was performed over three days using a structured spreadsheet documenting: (1) the full citation text, (2) the assigned primary and secondary category codes, (3) the reasoning for classification, and (4) key evidence from the GPTZero notes. This approach is consistent with established practices in research integrity content analysis where categories are based on observable features rather than subjective interpretation \cite{barbour2001checklists}.

%% file: results_revised.tex
\section{Results}

\subsection{Prevalence by Failure Mode}

Of the 100 hallucinated citations analyzed, we identified the following distribution across the five failure mode categories based on primary codes (Table \ref{tab:prevalence}):

\begin{table}[h]
\centering
\caption{Distribution of Hallucination Types in NeurIPS 2025 Papers (Primary Codes)}
\label{tab:prevalence}
\begin{tabular}{lcc}
\hline
\textbf{Category} & \textbf{Count} & \textbf{Percentage} \\
\hline
Total Fabrication (TF) & 66 & 66.0\% \\
Partial Attribute Corruption (PAC) & 27 & 27.0\% \\
Identifier Hijacking (IH) & 4 & 4.0\% \\
Placeholder Hallucination (PH) & 2 & 2.0\% \\
Semantic Hallucination (SH) & 1 & 1.0\% \\
\hline
\textbf{Total} & \textbf{100} & \textbf{100.0\%} \\
\hline
\end{tabular}
\end{table}

\begin{figure}[tb]
\centering
\includegraphics[width=0.85\textwidth]{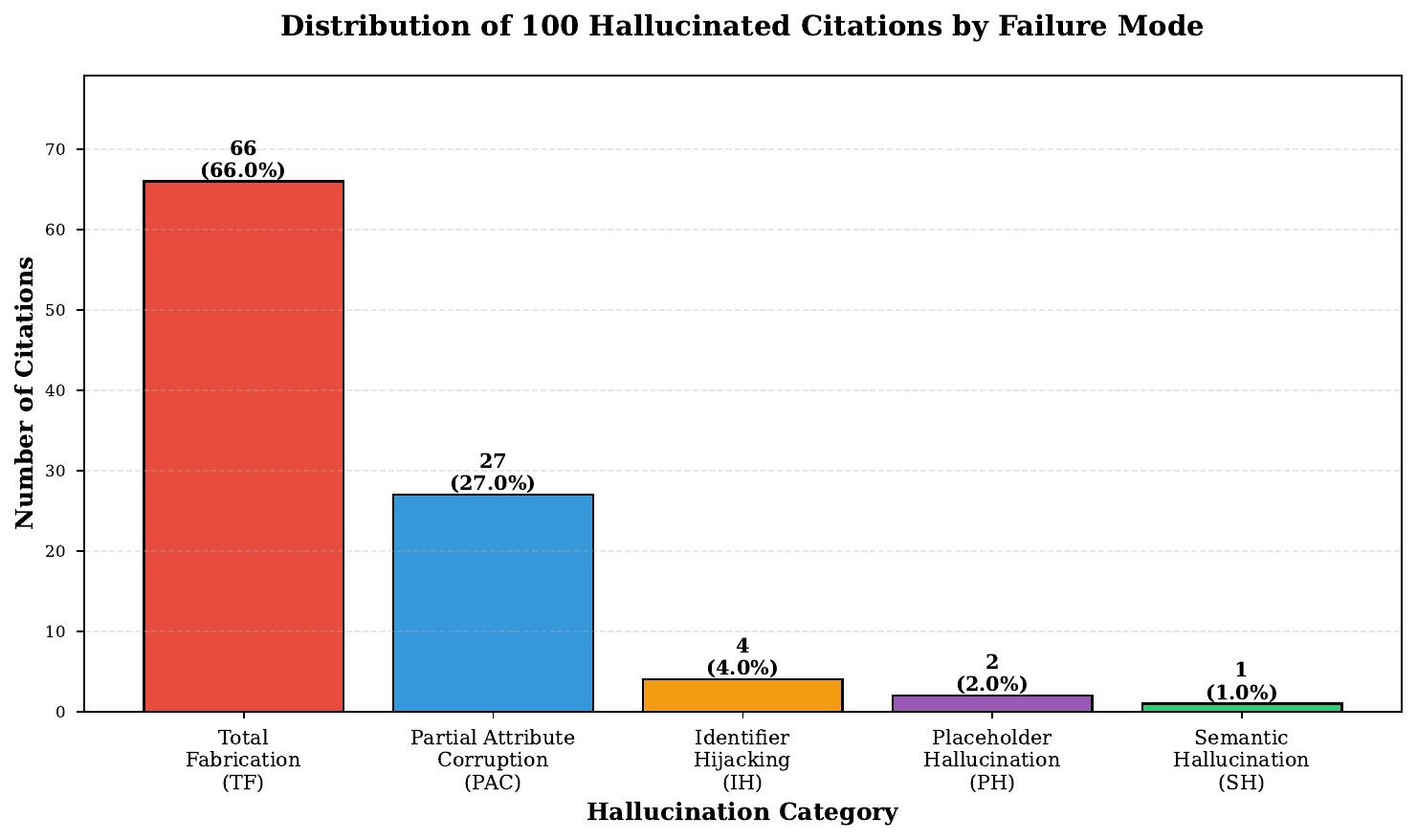}
\caption{Distribution of 100 hallucinated citations by primary failure mode. Total Fabrication dominates at 66\%, with Partial Attribute Corruption accounting for most remaining cases (27\%).}
\label{fig:distribution}
\end{figure}

The overwhelming prevalence of Total Fabrication (66\%) indicates that AI-generated hallucinations primarily involve wholesale invention rather than corruption of real scholarly metadata. Partial Attribute Corruption accounts for over one-quarter of cases (27\%), suggesting a secondary strategy of blending real and fabricated elements. The remaining categories: Identifier Hijacking (4\%), Placeholder Hallucination (2\%), and Semantic Hallucination (1\%), represent less common but potentially more sophisticated failure modes.

Table \ref{tab:examples} provides representative examples of each failure mode, illustrating the range of detection difficulty from trivial (placeholder text) to sophisticated (identifier hijacking with false verifiability). These examples demonstrate how hallucinations exploit different aspects of the citation verification process.

\begin{table*}[tb]
\centering
\caption{Representative Examples of Each Hallucination Failure Mode}
\label{tab:examples}
\small
\begin{tabular}{p{2.2cm}p{5.3cm}p{5.3cm}p{2.7cm}}
\toprule
\textbf{Category} & \textbf{Hallucinated Citation Example} & \textbf{Key Failure Indicator} & \textbf{Detection Difficulty} \\
\midrule

\textbf{Total Fabrication (TF)} & 
John Smith and Jane Doe. Deep learning techniques for avatar-based interaction in virtual environments. \textit{IEEE Transactions on Neural Networks and Learning Systems}, 32(12):5600--5612, 2021. &
All elements fabricated: generic author names (``John Smith and Jane Doe''), non-existent article, invalid DOI and URL. Zero overlap with any real publication. &
Easy (should be trivial) \\

\midrule

\textbf{Partial Attribute Corruption (PAC)} & 
Mario Paolone, Trevor Gaunt, Xavier Guillaud, Marco Liserre, Sakis Meliopoulos, Antonello Monti, Thierry Van Cutsem, Vijay Vittal, and Costas Vournas. A benchmark model for power system stability controls. \textit{IEEE Transactions on Power Systems}, 35(5):3627--3635, 2020. &
The authors match a real 2020 paper, but the title, publisher, volume, issue, and page numbers are all incorrect. Exploits author name recognition while corrupting bibliographic details. &
Hard (requires full metadata verification) \\

\midrule

\textbf{Identifier Hijacking (IH)} & 
Alex Wang, Rishi Bommasani, Dan Hendrycks, Daniel Song, and Zhilin Zhang. Efficient few-shot learning with EFL: A single transformer for all tasks. arXiv:2107.13586, 2021. &
Valid arXiv ID (2107.13586) links to real paper with completely different authors and title. Creates false verifiability---link works, content doesn't match. Exploits ``link works'' heuristic. &
Very Hard (false confirmation) \\

\midrule

\textbf{Semantic Hallucination (SH)} & 
Nanda, N. (2023). Progress in mechanistic interpretability: Reverse-engineering induction heads in GPT-2. &
Plausible-sounding title that fits the mechanistic interpretability domain conceptually. Author may be Neel Nanda, who wrote similar articles in 2023, but this specific title doesn't exist. &
Medium (sounds professional) \\

\midrule

\textbf{Placeholder Hallucination (PH)} & 
Firstname Lastname and Others. Drivlme: A large-scale multi-agent driving benchmark, 2023. URL or arXiv ID to be updated. &
Template variables not populated (``Firstname Lastname''). Obvious generation failure where LLM did not complete its task. Text like ``to be updated'' reveals incomplete generation. &
Trivial (immediately visible) \\

\bottomrule
\end{tabular}
\end{table*}

\subsection{Total Fabrication (TF): Complete Invention}

Total Fabrication represents citations with no correspondence to reality. These hallucinations invent authors, titles, venues, and identifiers entirely. Example:

\begin{quote}
\textit{John Smith and Jane Doe. Deep learning techniques for avatar-based interaction in virtual environments. IEEE Transactions on Neural Networks and Learning Systems, 32(12):5600--5612, 2021.}
\end{quote}

This citation uses obviously generic names (``John Smith and Jane Doe''), and verification reveals that no such article exists in the cited journal volume. The DOI and URL provided in the original citation were also fabricated. Despite the professional formatting and plausible venue, the entire citation is invented.

Another example demonstrates fabrication of specialized technical content:

\begin{quote}
\textit{Zhipeng Zhang, Chang Liu, Shihan Wu, and Yan Zhao. EST: Event spatio-temporal transformer for object recognition with event cameras. In ICASSP 2023 - 2023 IEEE International Conference on Acoustics, Speech and Signal Processing (ICASSP), pages 1-5. IEEE, 2023.}
\end{quote}

The title uses domain-appropriate terminology (``event spatio-temporal transformer,'' ``event cameras''), yet no matching paper exists in ICASSP 2023 proceedings. This pattern appeared repeatedly: technically coherent titles that sound appropriate for the research domain but correspond to no actual publication.

The 66\% prevalence of Total Fabrication indicates that AI language models default to wholesale invention when generating citations, creating complete fictional sources rather than attempting to corrupt or reference real scholarly work.

A third example involves apparent computer vision content:
\begin{quote}
	\textit{Z. Zhu, T. Yu, X. Zhang, J. Li, Y. Zhang, and Y. Fu. Neuralrgb-d: Neural representations for depth estimation and scene mapping. In Proceedings of the IEEE Conference on Computer Vision and Pattern Recognition, 2022.}
\end{quote}

The title uses appropriate terminology for CVPR, yet no matching paper exists in the 2022 proceedings. However, subsequent investigation revealed an important nuance. A 2022 version of another paper (Beltran et al., 2024, NFL-BA: Improving endoscopic SLAM with near-field light bundle adjustment, arXiv:2412.13176v1) contained this exact fabricated citation in its references, though it was corrected in later versions. This suggests the hallucination may not have originated with the NeurIPS author's LLM but was instead \textit{inherited} from contaminated training data. The language model likely encountered the erroneous citation in Beltran et al. (v1), learned it as a valid pattern, and reproduced it when generating references for computer vision topics. This mechanism represents a distinct failure mode: rather than fabricating citations \textit{de novo}, the model propagates pre-existing errors present in its training corpus. We have named this failure mode as ``Contamination Inheritance (CI).'' Such cases blur the boundary between AI hallucination and contamination-driven reproduction, underscoring that even `hallucinated' citations may have traceable origins in polluted datasets.

\subsection{Partial Attribute Corruption (PAC): Strategic Blending}

Partial Attribute Corruption citations combine real and fabricated elements, creating a hybrid that appears legitimate at first glance but fails detailed verification. A particularly sophisticated example:

\begin{quote}
\textit{Mario Paolone, Trevor Gaunt, Xavier Guillaud, Marco Liserre, Sakis Meliopoulos, Antonello Monti, Thierry Van Cutsem, Vijay Vittal, and Costas Vournas. A benchmark model for power system stability controls. IEEE Transactions on Power Systems, 35(5):3627--3635, 2020.}
\end{quote}

These authors did collaborate on a 2020 paper, but the title, volume, issue, and page numbers are all incorrect. A reviewer who verifies only the author list or publication year might miss the corrupted bibliographic details. This exploitation of partial recognition, where familiar names create a false sense of legitimacy, characterizes the PAC failure mode.

Another PAC pattern involves corrupted author lists on real papers:

\begin{quote}
\textit{Zayne Sprague, Xi Ye, Kyle Richardson, and Greg Durrett. MuSR: Testing the limits of chain-of-thought with multistep soft reasoning. In EMNLP, 2023.}
\end{quote}

The paper exists, but two authors are omitted and one (Kyle Richardson) is added. The paper was actually published at ICLR 2024, not EMNLP 2023. This type of corruption exploits the difficulty of verifying complete author lists and publication venues without consulting the actual source. Moreover, the inclusion of recognizable names from the research community may reduce reviewer scrutiny. When reviewers encounter familiar author names in their own domain, they are more likely to assume citation validity based on pattern recognition rather than perform granular verification of metadata.

\subsection{Identifier Hijacking (IH): False Verifiability}

Identifier Hijacking represents the most insidious failure mode. These citations provide valid scholarly identifiers (arXiv IDs, DOIs) that link to real papers, but the metadata does not match. An example:

\begin{quote}
\textit{Alex Wang, Rishi Bommasani, Dan Hendrycks, Daniel Song, and Zhilin Zhang. Efficient few-shot learning with EFL: A single transformer for all tasks. arXiv:2107.13586, 2021.}
\end{quote}

The arXiv identifier 2107.13586 is valid and links to a real paper. However, that paper has a completely different title and author list. A reviewer who clicks the link sees a working arXiv paper and may assume the citation is correct without verifying the content matches the claimed metadata.

This pattern creates false verifiability: the citation appears to pass a basic check (``the link works'') while failing substantive verification (``the content matches''). Of the 4 IH cases identified as primary failures, all involved valid identifiers pointing to unrelated papers, suggesting a systematic pattern in how language models generate or retrieve citation data.

\subsection{Placeholder Hallucination (PH): Generation Failures}

Placeholder Hallucinations reveal obvious AI generation failures where the model did not complete its task. Example:

\begin{quote}
\textit{Firstname Lastname and Others. Drivlme: A large-scale multi-agent driving benchmark, 2023. URL or arXiv ID to be updated.}
\end{quote}

The placeholder text ``Firstname Lastname'' is a clear indicator that the language model used template variables it failed to populate. The phrase ``URL or arXiv ID to be updated'' further confirms incomplete generation. Despite the obviousness of these errors, two such citations appeared in published NeurIPS papers, indicating that neither authors nor reviewers performed basic verification.

\subsection{Semantic Hallucination (SH): Plausible Fabrication}

Semantic Hallucinations invent conceptually appropriate titles that do not correspond to real papers. This was the rarest primary failure mode (1\% of cases) but represents sophisticated failures. Example:

\begin{quote}
\textit{Nanda, N. (2023). Progress in mechanistic interpretability: Reverse-engineering induction heads in GPT-2.}
\end{quote}

The title ``Progress in mechanistic interpretability: Reverse-engineering induction heads in GPT-2'' sounds professionally appropriate and the author may be Neel Nanda, a known researcher in mechanistic interpretability who published several related works in 2023. However, this specific title does not exist. The hallucination exploits domain knowledge to create a citation that fits the research context perfectly while corresponding to no actual publication.

\subsection{Distribution Across Papers}

The 100 hallucinations span 53 distinct NeurIPS papers, with a mean of 1.89 hallucinations per contaminated paper (range: 1--13, median: 2). The distribution exhibits a clear bimodal pattern: 49 of 53 papers (92\%) contain only 1--2 hallucinations, while 4 outlier papers contain 4--13 hallucinations each. This suggests two distinct author behaviors: cautious consultation of AI tools for ``polishing'' or ``gap-filling'' citations (the majority) versus systematic reliance on AI-generated citations throughout the writing process (a small minority).

The paper with the highest number of hallucinations (13 citations): ``Efficient semantic uncertainty quantification in language models via diversity-steered sampling'' demonstrates that citation fabrication can be extensive rather than isolated, suggesting reliance on AI tools throughout the writing process rather than occasional consultation.

\begin{figure}[htbp]
\centering
\includegraphics[width=0.8\textwidth]{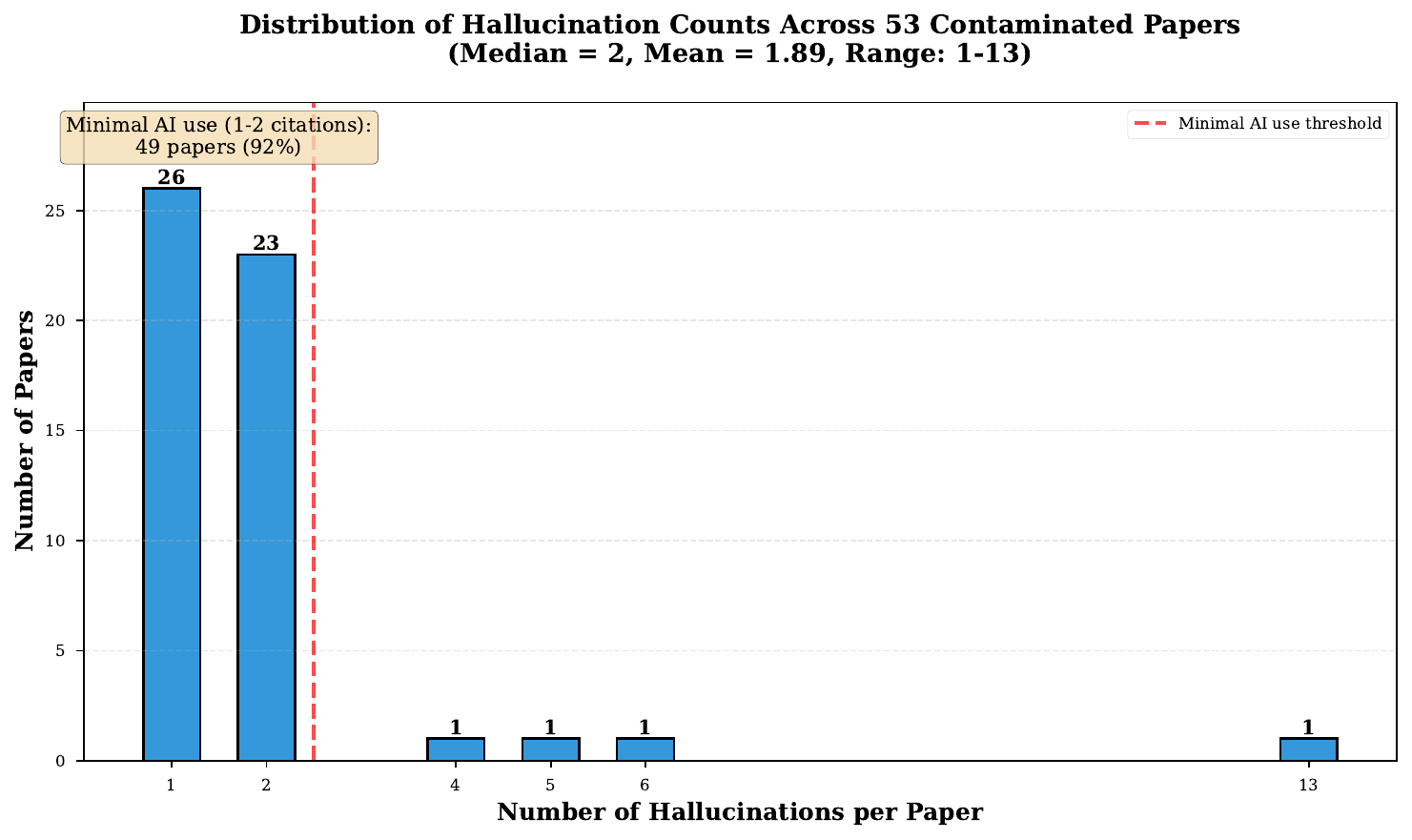}
\caption{Distribution of hallucination counts across 53 contaminated papers. The bimodal pattern shows 49 papers (92\%) with minimal AI use (1--2 citations), 3 papers (6\%) with heavy reliance (4--6 citations), and 1 outlier (2\%) with extensive reliance (13 citations). Median = 2, Mean = 1.89, Range: 1--13.}
\label{fig:histogram}
\end{figure}

\subsection{Compound Failure Modes: A Critical Finding}

A critical finding emerged from detailed analysis of secondary failure characteristics: every single hallucination in the dataset (100\%) exhibited compound failure modes, displaying characteristics of multiple deception mechanisms simultaneously. This is not simply methodological noise, it reveals that AI-generated hallucinations are fundamentally multi-layered fabrications rather than simple single-mechanism errors. Table \ref{tab:secondary} presents the distribution of secondary failure characteristics across all 100 citations.

\begin{table}[h]
\centering
\caption{Distribution of Secondary Failure Characteristics}
\label{tab:secondary}
\begin{tabular}{lcc}
\hline
\textbf{Secondary Category} & \textbf{Count} & \textbf{Percentage} \\
\hline
Semantic Hallucination (SH) & 63 & 63.0\% \\
Identifier Hijacking (IH) & 29 & 29.0\% \\
Placeholder Hallucination (PH) & 4 & 4.0\% \\
Partial Attribute Corruption (PAC) & 3 & 3.0\% \\
Total Fabrication (TF) & 1 & 1.0\% \\
\hline
\textbf{Total} & \textbf{100} & \textbf{100.0\%} \\
\hline
\end{tabular}
\end{table}

The dominant pattern involved Total Fabrications (66\% primary) that also employed Semantic Hallucination characteristics (63\% secondary across all citations). Specifically, among the 66 Total Fabrications, 50 cases (76\%) utilized Semantic Hallucination as their secondary mode to ensure the fabricated title appeared domain-appropriate. These citations invent all bibliographic metadata from scratch while using domain-appropriate terminology and plausible-sounding titles that fit the research context. For example, a fabricated citation might claim:

\begin{quote}
\textit{J. Smith and A. Patel. Leveraging large language models for financial forecasting. International Journal of Financial Technology, 9(2):101-115, 2024.}
\end{quote}

This is a completely invented source: no such authors, journal, or article exists. Yet the title ``Leveraging large language models for financial forecasting'' sounds professionally appropriate for an AI research paper, and the journal name ``International Journal of Financial Technology'' follows plausible naming conventions for academic venues. The semantic plausibility masks the total fabrication, making the citation appear legitimate at first glance.

Identifier Hijacking appeared as a secondary characteristic in 29\% of cases, typically combined with Total Fabrication or Partial Attribute Corruption as the primary failure mode. This suggests that false verifiability through working links is often layered onto other deception mechanisms rather than being used in isolation. For example:

\begin{quote}
\textit{Mehdi Azabou, Micah Weber, Wenlin Ma, et al. Mineclip: Multimodal neural exploration of clip latents for automatic video annotation. arXiv preprint arXiv:2210.02870, 2022.}
\end{quote}

The arXiv ID 2210.02870 is valid and links to a real paper, but that paper has completely different authors and title. The citation combines Total Fabrication (invented metadata) with Identifier Hijacking (valid but mismatched link), creating a compound deception that exploits both semantic plausibility and false verifiability.

This 100\% compound failure rate has profound implications for detection strategies. Simple verification of any single attribute (author names, title keywords, identifier validity) will fail to catch these hallucinations because each component may appear partially legitimate. Only comprehensive cross-verification, confirming that authors, title, venue, date, and identifier all match the same actual source, can reliably detect AI-generated citations.

The strategic layering of deception mechanisms reveals that while LLMs may not be `intentionally' deceptive, the statistical patterns they learn from training data lead them to generate citations that systematically exploit multiple verification heuristics simultaneously. The fact that 76\% of Total Fabrications specifically layer semantic plausibility on top of wholesale invention demonstrates this is not random noise but structured fabrication that anticipates how humans verify citations superficially rather than comprehensively.

%% file: discussion_revised.tex
\section{Discussion}

\subsection{The Systematic Failure of Elite Peer Review}

Each paper accepted by NeurIPS 2025 underwent review by 3--5 expert researchers, yet 53 papers containing 100 fabricated citations were published. This represents approximately 1\% contamination rate at one of the world's most selective AI conferences. The failure is systematic, not isolated: hallucinated citations appeared across multiple institutions, research areas, and paper types, indicating that current peer review processes do not include effective citation verification. The observable facts are these: (1) Multiple expert reviewers examined each paper; (2) Two citations contained obvious errors (``Firstname Lastname,'' ``URL or arXiv ID to be updated''); (3) None were caught during review. This indicates that citation checking is not part of the standard reviewer workflow at elite conferences, even when reviewing papers about AI systems known to hallucinate.

The volume problem exacerbates this failure. NeurIPS 2025 received 21,575 submissions, requiring thousands of volunteer reviewers.This submission tsunami reflects broader trends in AI-driven content proliferation, where automation enables mass production of superficially competent but substantively hollow outputs across academic, corporate, and creative domains \cite{ansari2025slop}.  Under these conditions, detailed verification of every citation becomes practically impossible. However, this does not excuse the failure, it reveals that the peer review system has scaled beyond its verification capacity. 

\subsection{Why Compound Failures Succeed: The Multi-Layered Deception Strategy}

Our analysis reveals that AI-generated hallucinations employ compound deception strategies rather than single failure mechanisms. Every hallucination in our dataset exhibited characteristics of multiple failure modes simultaneously, with the most common pattern being Total Fabrication (66\% primary) layered with Semantic Hallucination (63\% secondary).

This compound structure explains why peer review fails to detect these fabrications. Consider the verification heuristics that a time-pressed reviewer might employ:

\textbf{Heuristic 1: ``Does the title sound appropriate for this research area?''} The 63\% secondary Semantic Hallucination rate across all citations (with 76\% of Total Fabrications specifically employing this strategy) indicates that most fabricated citations use domain-appropriate terminology and plausible-sounding titles. A citation claiming ``Leveraging large language models for financial forecasting'' sounds professionally legitimate for an AI paper, passing this superficial check even though the entire citation is invented.

\textbf{Heuristic 2: ``Does the link/identifier work?''} The 29\% secondary Identifier Hijacking rate across all citations shows that many fabrications include valid arXiv IDs or DOIs that link to real papers. A reviewer who clicks the link sees an actual scholarly article and may assume correctness without verifying that the metadata matches. This creates false verifiability i.e. the citation passes the ``link works'' test while failing comprehensive verification.

\textbf{Heuristic 3: ``Do I recognize these author names or venues?''} Partial Attribute Corruption (27\% primary) exploits name recognition by attributing real scholars to papers they didn't write, or citing real papers with corrupted metadata. Reviewers may recognize familiar names (Kingma, Ba, Bengio, Durrett) or venue abbreviations (ICLR, NeurIPS, CVPR) without checking that the specific combination is correct.

\begin{figure}[htbp]
\centering
\includegraphics[width=0.90\textwidth]{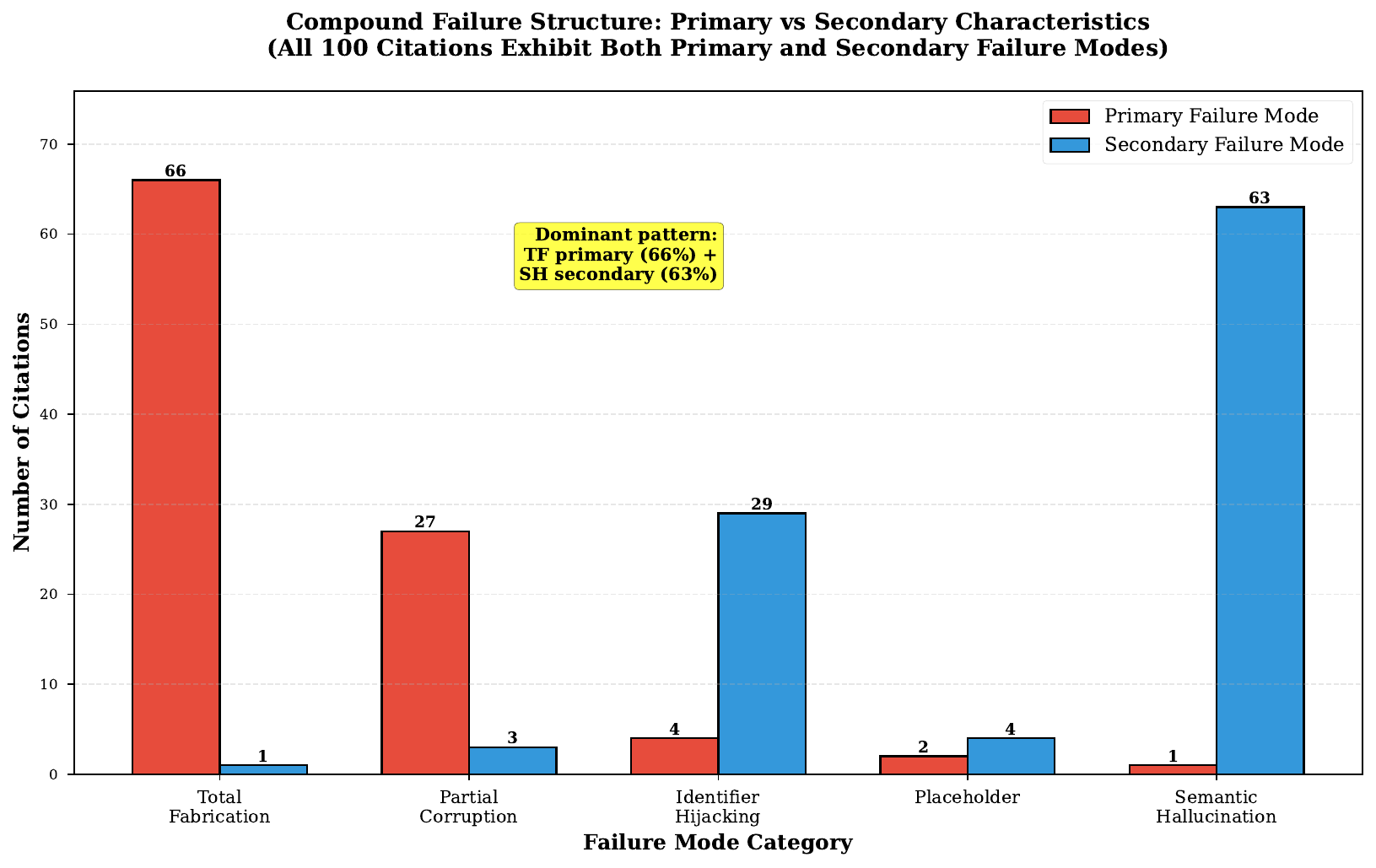}
\caption{Compound failure structure of AI-generated hallucinations. The Venn diagram shows overlap between primary failure modes (outer circles) and secondary characteristics (inner regions). The dominant pattern (TF primary + SH secondary, 63\% of all citations) combines wholesale fabrication with semantic plausibility, exploiting the ``sounds right'' heuristic while providing zero verifiable content. IH secondary (29\%) creates false verifiability by providing working links to unrelated papers.}
\label{fig:compound}
\end{figure}

The compound failure structure means that citations designed to pass multiple superficial checks simultaneously will evade detection unless reviewers perform comprehensive cross-verification, confirming that authors, title, venue, date, and identifier all match the same actual source. Our data suggests this level of verification is not standard practice in peer review. Figure \ref{fig:compound} illustrates how compound failures create overlapping deception layers that must all be checked to detect fabrication.

\subsection{The Dominance of Total Fabrication: Implications for Detection}

The 66\% prevalence of Total Fabrication as the primary failure mode has important implications for understanding how LLMs generate citations and how to detect them.

\textbf{LLMs default to invention, not corruption.} One might hypothesize that language models trained on scholarly literature would generate citations by retrieving and corrupting real sources. Our data contradicts this: two-thirds of hallucinations involve wholesale invention with zero overlap with reality. This suggests that LLMs generate citations through the same next-token prediction mechanism that produces all text, rather than through retrieval and modification of real bibliographic data.

\textbf{Detection can exploit the fabrication baseline.} Because 66\% of hallucinations are Total Fabrications, automated verification tools that simply check ``does this exact citation exist?'' will catch the majority of cases. The current GPTZero approach: web search plus academic database lookup, is well-matched to this threat model.

\textbf{The 27\% PAC cases require more sophisticated verification.} Partial Attribute Corruption involves blending real and fake elements, requiring verification tools to check not just existence but metadata consistency. For example, verifying that the claimed authors actually wrote the paper with the claimed title, published in the claimed venue, in the claimed year. This is computationally more expensive but necessary to catch the remaining quarter of hallucinations.

\textbf{Compound failures (100\% rate) mean single-attribute checking fails.} Even though 66\% are Total Fabrications, all of them layer semantic plausibility on top of the invention. A tool that only checks ``does this title sound plausible?'' or ``do these author names exist?'' will generate false negatives. Comprehensive verification requires checking multiple attributes and confirming they all point to the same real source.

\subsection{Addressing the ``Harmless Error'' Counterargument}

Some may argue that citation errors are minor formatting issues that do not invalidate research findings. This view fundamentally misunderstands the epistemic function of citations in scientific discourse.

Citations serve three critical functions: (1) \textit{Evidentiary support:} they establish that claimed prior work actually exists and says what the authors claim; (2) \textit{Reproducibility:} they enable readers to trace findings back to source data and methods; (3) \textit{Credit attribution:} they acknowledge intellectual debt and situate new work within existing knowledge.

Fabricated citations subvert all three functions. They make false evidentiary claims (``This result has been established by prior work'' when no such work exists). They break reproducibility (readers cannot verify or build upon non-existent sources). They misallocate credit (attributing work to authors who did not produce it, as in the Paolone et al. example where real authors are credited with a paper they never wrote).

This is not analogous to a typographical error in a correctly cited source. A typo writes ``Smtih'' instead of ``Smith''$\rightarrow$ the reference is identifiable and verifiable. A hallucination fabricates ``Smith, J. et al. (2024). Paper that does not exist'' $\rightarrow$ the reference is fundamentally fraudulent, whether or not the fraud was intentional.

Moreover, the ``core math is sound'' defense ignores how scientific knowledge accumulates. Research builds on prior work through citation linkages. If those linkages are fabricated, the entire knowledge graph becomes corrupted. Future researchers who trust these citations will waste time searching for non-existent sources or will propagate the hallucinations in their own work.

The 100\% compound failure rate makes this problem worse. These are not simple mistakes, they are multi-layered fabrications that actively mislead readers by appearing legitimate through semantic plausibility, familiar author names, or working identifiers that point to different papers.

\subsection{Beyond NeurIPS: A Systemic Problem}

The NeurIPS contamination is not unique. GPTZero's analysis of ICLR 2026 submissions identified over 50 additional hallucinations in papers under review, with some papers receiving average ratings of 8/10 despite containing fabricated citations \cite{gptzero2025iclr}. This suggests that absent intervention, contaminated papers would have been accepted at ICLR as they were at NeurIPS.

The U.S. government's ``Make America Healthy Again'' report contained fabricated citations that were identified only after publication \cite{weber_maha_2025}. Deloitte Australia was forced to refund (AUD)\$98,000 after fabricated citations were discovered in a government report they produced \cite{gptzero2025deloitte}.

These cases span academic conferences, government agencies, and professional consulting firms, all contexts with ostensibly rigorous quality control. The common factor is reliance on human verification in an environment where AI tools can generate thousands of plausible-looking but fabricated citations faster than humans can verify them. This dynamic exemplifies what has been characterized as a recursive contamination cycle, wherein synthetic outputs saturate information ecosystems and undermine the verification infrastructure upon which knowledge systems depend \cite{ansari2025slop}.

The problem is not isolated to low-quality venues or negligent authors. It appears at the most prestigious conferences in AI, produced by authors at elite institutions (NYU, Cambridge, MIT, Genentech, Google). This suggests the contamination is driven by systemic factors: tool accessibility, publication pressure, inadequate verification infrastructure, rather than individual misconduct.

\subsection{Solutions: Mandatory Automated Verification}

The solution is straightforward in principle: require automated citation verification at the point of submission. Tools already exist for this purpose, including GPTZero's Hallucination Check, Crossref's citation checking APIs, and academic database verification protocols.

Conference organizers should implement two key measures. First, mandate citation verification as part of the submission process, similar to how plagiarism detection is now standard. Provide authors with verification reports before peer review begins, allowing them to correct errors. Second, flag papers with unverifiable citations for enhanced reviewer scrutiny and train reviewers in citation verification techniques.

The compound failure structure we documented suggests that verification tools must check multiple attributes simultaneously and confirm they all reference the same source:

\begin{enumerate}
\item \textbf{Existence check}: Does this exact citation exist in web search or academic databases? (Catches 66\% TF primary cases)
\item \textbf{Metadata consistency check}: Do the claimed authors, title, venue, and date all match the same real source? (Catches 27\% PAC cases)
\item \textbf{Identifier validation check}: If an arXiv ID or DOI is provided, does the paper at that identifier match the claimed metadata? (Catches 4\% IH primary and 29\% IH secondary cases)
\item \textbf{Semantic plausibility check}: Does the title use domain-appropriate terminology? (Useful for flagging suspicious citations for human review, though 63\% of fabrications pass this check)
\end{enumerate}

Requiring all four checks would catch the vast majority of hallucinations in the dataset. The technical barriers are minimal, these tools are already deployed in research integrity investigations. The primary barrier is institutional: conference organizers must acknowledge that the current system has failed and implement verification requirements despite potential author pushback.

The alternative is normalization. If approx. 1\% contamination at NeurIPS becomes ``acceptable,'' the problem will worsen as AI tools become more sophisticated at generating plausible-looking fabrications. The 2026 conference cycle represents an opportunity to prevent this failure mode from becoming endemic.

\subsection{Limitations}

Our analysis has several limitations. First, we analyzed only citations flagged by GPTZero's automated tool. Citations that appear correct but subtly misrepresent source content would not be detected. Second, we cannot determine author \textit{intent}, whether hallucinations resulted from deliberate fraud, negligent use of AI tools, or honest mistakes. Our analysis focuses on the observable failure (citations are fabricated) and the systemic gap (reviewers did not catch them), not on attributing blame.

Third, our taxonomy classifies hallucinations by their most prominent failure characteristic as the primary code, with secondary characteristics captured separately. The boundaries between categories (particularly PAC vs. TF) can be ambiguous when citations blend multiple real and fake elements. However, the 100\% compound failure rate we documented suggests this ambiguity reflects the fundamental nature of AI-generated hallucinations rather than coding imprecision. Finally, the data comes from a single conference (NeurIPS 2025) in a single discipline (AI/ML). While parallel findings at ICLR, in government reports, and in consulting outputs suggest broader applicability, we cannot claim that our 66\% TF / 27\% PAC distribution \textit{generalizes} to all domains or all AI writing tools. Different LLMs, different prompting strategies, or different research fields might produce different hallucination patterns.

\subsection{Future Work}

\subsubsection{On Contamination Inheritance}

Our five-category taxonomy (TF, PAC, IH, PH, SH) was designed to classify hallucinations based on observable output characteristics rather than generative mechanisms. However, the discovery of the Contamination Inheritance failure mode, wherein LLMs reproduce pre-existing erroneous citations from contaminated training data rather than fabricating them \textit{de novo}, suggests that additional classification dimensions may be necessary to fully understand AI-generated citation failures.

Contamination Inheritance was identified in at least one case in our dataset: the Z. Zhu et al. citation discussed in Section 3.2, which traced back to an earlier arXiv preprint that was subsequently corrected. This finding raises several questions that warrant deeper investigation:

\begin{itemize}
	\item What proportion of citations we classified as Total Fabrication are actually Contamination Inheritances from contaminated training sources?
	\item Can training-data provenance tools distinguish between pure fabrication and learned reproduction?
	\item Do Contamination Inheritances exhibit different detection profiles than \textit{de novo} fabrications, potentially requiring specialized verification strategies?
\end{itemize}

Addressing these questions requires capabilities beyond output-level analysis. Future work should integrate training-data audits, version-controlled corpus tracking, and citation genealogy mapping to trace the origins of hallucinated citations. Such investigations would reveal whether the contamination feedback loop, where AI-generated errors are learned by subsequent models, is already widespread or represents isolated cases. This distinction has significant governance implications: if Contamination Inheritance dominates, mitigation strategies must focus on cleaning training corpora and implementing temporal watermarking to identify synthetic versus human-authored sources. If pure fabrication dominates, efforts should prioritize improving model grounding and retrieval-augmented generation architectures.

\subsubsection{On Other Improvements}

Additionally, the compound failure structure we documented (100\% of citations exhibiting both primary and secondary characteristics) suggests that future taxonomies should capture multi-dimensional failure patterns more systematically. Rather than assigning a single primary code, a matrix-based classification system could represent the simultaneous presence of fabrication, semantic plausibility, identifier hijacking, and error inheritance. Such a framework would better reflect the strategic, multi-layered nature of AI-generated hallucinations and inform more robust detection algorithms.

Finally, longitudinal studies tracking hallucination patterns across model generations, training datasets, and research domains would illuminate whether fabrication strategies are evolving in response to detection tools. If LLMs increasingly adopt Contamination Inheritance or compound strategies to evade verification, the arms race between generation and detection will require adaptive governance mechanisms that anticipate rather than react to emerging failure modes.

%% file: conclusion_revised.tex
\section{Conclusion}

This study documents 100 AI-generated hallucinated citations that appeared in papers accepted by NeurIPS 2025, one of the world's most prestigious AI conferences. Each paper was reviewed by 3--5 expert researchers, yet fabricated citations evaded detection throughout the peer review process. Our failure mode taxonomy reveals that these hallucinations are predominantly Total Fabrications (66\%), with citations invented wholesale rather than corrupted from real sources. However, a critical finding emerged: every single hallucination exhibited compound failure modes, with 63\% layering semantic plausibility onto fabricated content and 29\% incorporating identifier hijacking to create false verifiability. The appearance of fabricated citations in elite peer-reviewed venues is not a hypothetical future risk, it is a documented present reality. The problem extends beyond NeurIPS to other major conferences, government reports, and professional consulting outputs. The common factor is the use of AI writing tools without adequate verification infrastructure to detect generated fabrications. Three findings warrant immediate attention from conference organizers and research integrity specialists:

First, current peer review does not include systematic citation verification. Reviewers focus on methodological rigor, novelty, and experimental results, not on verifying that cited sources actually exist. This gap has always existed but becomes critical when AI tools can generate thousands of plausible-looking fabrications faster than humans can verify them. The two Placeholder Hallucinations in our dataset (``Firstname Lastname'' citations) demonstrate that even trivially detectable errors pass through review unnoticed.

Second, compound failure modes explain why hallucinations evade detection. The 100\% compound failure rate we documented reveals that AI-generated citations are not simple errors but multi-layered fabrications that exploit multiple verification heuristics simultaneously. A citation that sounds semantically plausible, includes a working link, and references familiar author names will pass superficial inspection even if comprehensive cross-verification would reveal it as completely fabricated. Detection requires checking multiple attributes and confirming they all point to the same real source.

Third, the solution is implementable now. Automated citation verification tools exist and are already used in research integrity investigations. However, because 100\% of hallucinations in our dataset exhibited compound failure modes, simple link-checking is insufficient. Verification tools must cross-reference authors, titles, venues, dates, and identifiers simultaneously to be effective. Conference organizers should mandate four-step verification: (1) existence check via web search and academic databases, (2) metadata consistency check confirming authors/title/venue match, (3) identifier validation confirming DOIs/arXiv IDs point to claimed papers, and (4) flagging semantically suspicious citations for human review. This multi-attribute verification strategy, matched to the compound failure structure we documented, would catch the vast majority of hallucinations.

The primary obstacle is institutional inertia, the reluctance to acknowledge that the current system has failed and implement verification requirements despite potential author pushback. The 2026 conference submission cycle represents a critical window. If mandatory verification is implemented before next year's major AI conferences (NeurIPS, ICLR, ICML, AAAI), it could prevent fabricated citations from becoming normalized in scientific literature. If organizers wait for ``more evidence'' or defer to author convenience, the 1\% contamination rate documented here will likely increase as AI tools become more sophisticated at generating plausible fabrications.

The irony is stark: the AI research community, which created the tools that generate these hallucinations, failed to detect them in its own peer review process. This failure is not an indictment of individual reviewers or authors, it is evidence that institutional verification infrastructure has not kept pace with AI capabilities. The question is whether the community will implement the solutions that already exist or continue relying on a peer review system that was never designed to catch AI-generated fabrications at scale.

The 100\% compound failure rate we documented reveals the fundamental challenge: AI-generated hallucinations are not random errors but structured fabrications that anticipate human verification behavior. They layer semantic plausibility, familiar names, and working links to create citations that look legitimate at every level of superficial inspection. Only comprehensive, automated verification can reliably detect them. The tools exist. The evidence is documented. The 2026 conference cycle is the opportunity to act.

%% file: references_updated.bib
@misc{gptzero2026neurips,
  author = {{GPTZero}},
  title = {{GPTZero} finds 100 new hallucinations in {NeurIPS} 2025 accepted papers},
  year = {2026},
  howpublished = {\url{https://gptzero.me/news/neurips/}},
  note = {Accessed: 2026-02-05}
}

@misc{gptzero2025iclr,
  author = {{GPTZero}},
  title = {{GPTZero} finds over 50 new hallucinations in {ICLR} 2026 submissions},
  year = {2025},
  howpublished = {\url{https://gptzero.me/news/iclr-2026/}},
  note = {Accessed: 2026-02-05}
}

@misc{gptzero2025deloitte,
  author = {{GPTZero}},
  title = {Deloitte's Citation Situation \& {GPTZero's} Citation Solution},
  year = {2025},
  howpublished = {\url{https://gptzero.me/news/deloitte-australia-citation-check/}},
  note = {Accessed: 2026-01-26}
}

@article{ji2023hallucination,
  title={Survey of hallucination in natural language generation},
  author={Ji, Ziwei and Lee, Nayeon and Frieske, Rita and Yu, Tiezheng and Su, Dan and Xu, Yan and Ishii, Etsuko and Bang, Ye Jin and Madotto, Andrea and Fung, Pascale},
  journal={ACM Computing Surveys},
  volume={55},
  number={12},
  pages={1--38},
  year={2023},
  publisher={ACM}
}

@article{huang2025survey,
  title={A survey on hallucination in large language models: Principles, taxonomy, challenges, and open questions},
author={Huang, Lei and Yu, Weijiang and Ma, Weitao and Zhong, Weihong and Feng, Zhangyin and Wang, Haotian and Chen, Qianglong and Peng, Weihua and Feng, Xiaocheng and Qin, Bing and others},
journal={ACM Transactions on Information Systems},
volume={43},
number={2},
pages={1--55},
year={2025},
publisher={ACM New York, NY}
}

@article{alkaissi2023artificial,
  title={Artificial hallucinations in {ChatGPT}: implications in scientific writing},
  author={Alkaissi, Hanan and McFarlane, Samy I},
  journal={Cureus},
  volume={15},
  number={2},
  year={2023},
  publisher={Cureus}
}

@article{lee2023benefits,
  title={Benefits, limits, and risks of {GPT-4} as an {AI} chatbot for medicine},
  author={Lee, Peter and Bubeck, Sebastien and Petro, Joseph},
  journal={New England Journal of Medicine},
  volume={388},
  number={13},
  pages={1233--1239},
  year={2023},
  publisher={Mass Medical Soc}
}

@article{dahl2024large,
  title={Large legal fictions: Profiling legal hallucinations in large language models},
  author={Dahl, Matthew and Magesh, Varun and Suzgun, Mirac and Ho, Daniel E},
  journal={Journal of Legal Analysis},
  volume={16},
  number={1},
  pages={64--93},
  year={2024}
}

@article{rawte2023survey,
  title={A survey of hallucination in large foundation models},
  author={Rawte, Vipula and Sheth, Amit and Das, Amitava},
  journal={arXiv preprint arXiv:2309.05922},
  year={2023}
}

@article{rivkin2020manuscript,
	title={Manuscript referencing errors and their impact on shaping current evidence},
	author={Rivkin, Anastasia},
	journal={American Journal of Pharmaceutical Education},
	volume={84},
	number={7},
	year={2020},
	publisher={Elsevier}
}

@article{fang2011misconduct,
  title={Misconduct accounts for the majority of retracted scientific publications},
author={Fang, Ferric C and Steen, R Grant and Casadevall, Arturo},
journal={Proceedings of the National Academy of Sciences},
volume={109},
number={42},
pages={17028--17033},
year={2012},
publisher={National Academy of Sciences}
}

@article{barbour2001checklists,
  title={Checklists for improving rigour in qualitative research: a case of the tail wagging the dog?},
  author={Barbour, Rosaline S},
  journal={BMJ},
  volume={322},
  number={7294},
  pages={1115--1117},
  year={2001},
  publisher={British Medical Journal Publishing Group}
}

@misc{weber_maha_2025,
	author       = {Lauren Weber and Caitlin Gilbert},
	title        = {White House {MAHA} Report May Have Garbled Science by Using {AI}, Experts Say},
	journal      = {The Washington Post},
	year         = {2025},
	month        = may,
	day          = {29},
	howpublished =  {\url{https://www.washingtonpost.com/health/2025/05/29/maha-rfk-jr-ai-garble/}},
	  note = {Accessed: 2026-02-05}
}

@article{ansari2025slop,
	author = {Samar Ansari},
	title = {{AI} Slop and Data Pollution in the Age of Generative {AI}: Strategic Risks, Economic Consequences, and Governance Pathways for Business, Management, and the Creative Industries},
	year = {2025},
	journal = {Available at SSRN, DOI: 10.2139/ssrn.5649410},
	doi = {10.2139/ssrn.5649410},
	howpublished = {\url{https://dx.doi.org/10.2139/ssrn.5649410}}
}
